\documentclass[a4paper]{jpconf}
\usepackage{graphicx}
\begin{document}

\def\be{\begin{equation}}
\def\ee{\end{equation}}
\def\ba{\begin{eqnarray}}
\def\ea{\end{eqnarray}}
\newcommand{\intdP}{\int\!dP}

\title{Quasiparticle anisotropic hydrodynamics }

\author{Mubarak Alqahtani and Michael Strickland}

\address{Kent State University, Kent, OH 44242 USA }

\ead{malqaht3@kent.edu (presenter of the talk)}

\begin{abstract}
We study an azimuthally-symmetric boost-invariant quark-gluon plasma using quasiparticle anisotropic hydrodynamics including the effects of both shear and bulk viscosities. We compare results obtained using the quasiparticle method with the standard anisotropic hydrodynamics and viscous hydrodynamics. We consider the predictions of the three methods for the differential particle spectra and  mean transverse momentum. We find that the three methods agree for small shear viscosity to entropy density ratio, $\eta/s$, but show differences at large $\eta/s$. Additionally, we find that the standard
anisotropic hydrodynamics method shows suppressed production at low transverse-momentum compared to the other two methods, and the bulk-viscous correction can drive the primordial particle spectra negative at large $ p_T$  in viscous hydrodynamics.
\end{abstract}

\section{Introduction}

Ultrarelativistic heavy-ion collision experiments at RHIC and LHC study the quark-gluon plasma (QGP) using high-energy nuclear collisions. Relativistic hydrodynamics has been quite successful in describing  the collective behavior seen in these experiments. 
Over the years, many approaches have been developed to model the physics of the QGP such as ideal hydrodynamics \cite{Huovinen:2001cy} and viscous hydrodynamics \cite{Romatschke:2007mq,Ryu:2015vwa,Niemi:2011ix}. Viscous hydrodynamic studies indicate that the QGP is a momentum-space anisotropic plasma. In order to more accurately account for potentially large momentum-space anisotropies, anisotropic hydrodynamics was developed \cite{Florkowski:2010cf,Martinez:2010sc,Martinez:2012tu,Ryblewski:2012rr,Bazow:2013ifa,Nopoush:2014pfa,Nopoush:2014qba}. In recent years, there have been developments in the anisotropic hydrodynamics program which focused on making phenomenological predictions for heavy-ion physics \cite{Alqahtani:2015qja,Alqahtani:2016rth,Nopoush:2015yga}. For a recent review on anisotropic hydrodynamics, see Ref. \cite{Strickland:2014pga}.

A major conceptual difficulty for anisotropic hydrodynamics is how to impose a realistic equation of state (EoS) on a system which is potentially far from equilibrium. In quasiparticle hydrodynamics (aHydroQP), one  imposes a realistic EoS which takes into account the non-conformality of QGP by introducing a single finite-temperature quasiparticle mass which is fit to lattice data. We compare aHydroQP with the standard anisotropic hydrodynamics (aHydro) where  approximate conformality is assumed. Additionally, we compare both aHydro formulations with second-order viscous hydrodynamics (vHydro) \cite{Denicol:2012cn,Denicol:2014vaa}. We then present comparisons of the primordial particle spectra and the average transverse momentum  for pions, kaons, and protons predicted by these three methods  \cite{Alqahtani:2016rth}. 

\section{Anisotropic hydrodynamics}
In non-conformal anisotropic hydrodynamics, the anisotropy tensor is defined as \cite{Nopoush:2014pfa}
\be
\Xi^{\mu\nu} = u^\mu u^\nu + \xi^{\mu\nu} - \Delta^{\mu\nu} \Phi \, ,
\ee
where $u^\mu$ is the fluid four-velocity, $\xi^{\mu\nu}$ is a symmetric and traceless anisotropy tensor, and $\Phi$ is associated with the bulk degree of freedom.
The one-particle distribution function is assumed to be of the form
\be
f(x,p) = f_{\rm iso}\!\left(\frac{1}{\lambda}\sqrt{p_\mu \Xi^{\mu\nu} p_\nu}\right) ,
\label{eq:genf}
\ee
where $ \lambda $ can be identified with the temperature only in the isotropic equilibrium limit. By expanding the argument of the square root in  Eq.~(\ref{eq:genf}), one can write
\be
f(x,p) =  f_{\rm eq}\!\left(\frac{1}{\lambda}\sqrt{\sum_i \frac{p_i^2}{\alpha_i^2} + m^2}\right) ,
\label{eq:fform}
\ee
where $i\in \{x,y,z\}$ and the ellipticity parameters $\alpha_i$ are $\alpha_i \equiv (1 + \xi_i + \Phi)^{-1/2} \,$.

\section{Equation of state}
In quasiparticle hydrodynamics, we implement a realistic equation of state by assuming that we have an ensemble of massive quasiparticles with a single temperature-dependent mass. However, in order to not violate thermodynamical consistency, one must introduce a background field contribution to the energy-momentum tensor \cite{Alqahtani:2015qja,Romatschke:2011qp}
\be
T^{\mu\nu}_{\rm eq} = T^{\mu\nu}_{\rm kinetic,eq} + g^{\mu\nu} B_{\rm eq}  \, ,
\ee
where $B_{\rm eq}$ can be found by integrating 
\ba 
\frac{dB_{\rm eq}}{dT} 
&=& -4\pi \tilde{N}m^2 T K_1(\hat{m}_{\rm eq}) \frac{dm}{dT} \, ,
\ea
with $ m(T)$ obtained from this thermodynamic identity
\be
{\cal E}_{\rm eq}+{\cal P}_{\rm eq}=T{\cal S}_{\rm eq} = 4 \pi \tilde{N} T^4 \, \hat{m}_{\rm eq}^3 K_3\left( \hat{m}_{\rm eq}\right) ,
\label{eq:meq}
\ee
where $ {\cal E}_{\rm eq}$ and $ {\cal P}_{\rm eq}$ are provided from lattice QCD calculations \cite{Borsanyi:2010cj}.

\section{Boltzmann equation}
The dynamical equations necessary to describe the system can be obtained by taking moments of Boltzmann equation \cite{Alqahtani:2015qja,Romatschke:2011qp}
\be
p^\mu \partial_\mu f+\frac{1}{2}\partial_i m^2\partial^i_{(p)} f=-\mathcal{C}[f]\,,
\label{eq:boltz2}
\ee
where $ C[f] $ is the collisional kernel containing all interactions involved. For this work, we specialize to the relaxation time approximation (RTA) for the collisional kernel.

\section{Anisotropic freeze-out}
The freeze-out process occurs at the late times, leading to the particle production observed experimentally. In anisotropic hydrodynamics, we perform  ``anisotropic Cooper-Frye freeze-out'' using the distribution function specified in  Eq.~(\ref{eq:genf}), which ensures that the distribution function is positive definite in all regions in phase space \cite{Alqahtani:2016rth,Nopoush:2015yga}. In viscous hydrodynamics, however, one linearizes around the equilibrium distribution function to take into account the dissipative correction and the positivity of the distribution function is no longer guaranteed \cite{Rose:2014fba,Bozek:2009dw}.

\begin{figure}[t]
\hspace{5cm}
\includegraphics[width=13pc]{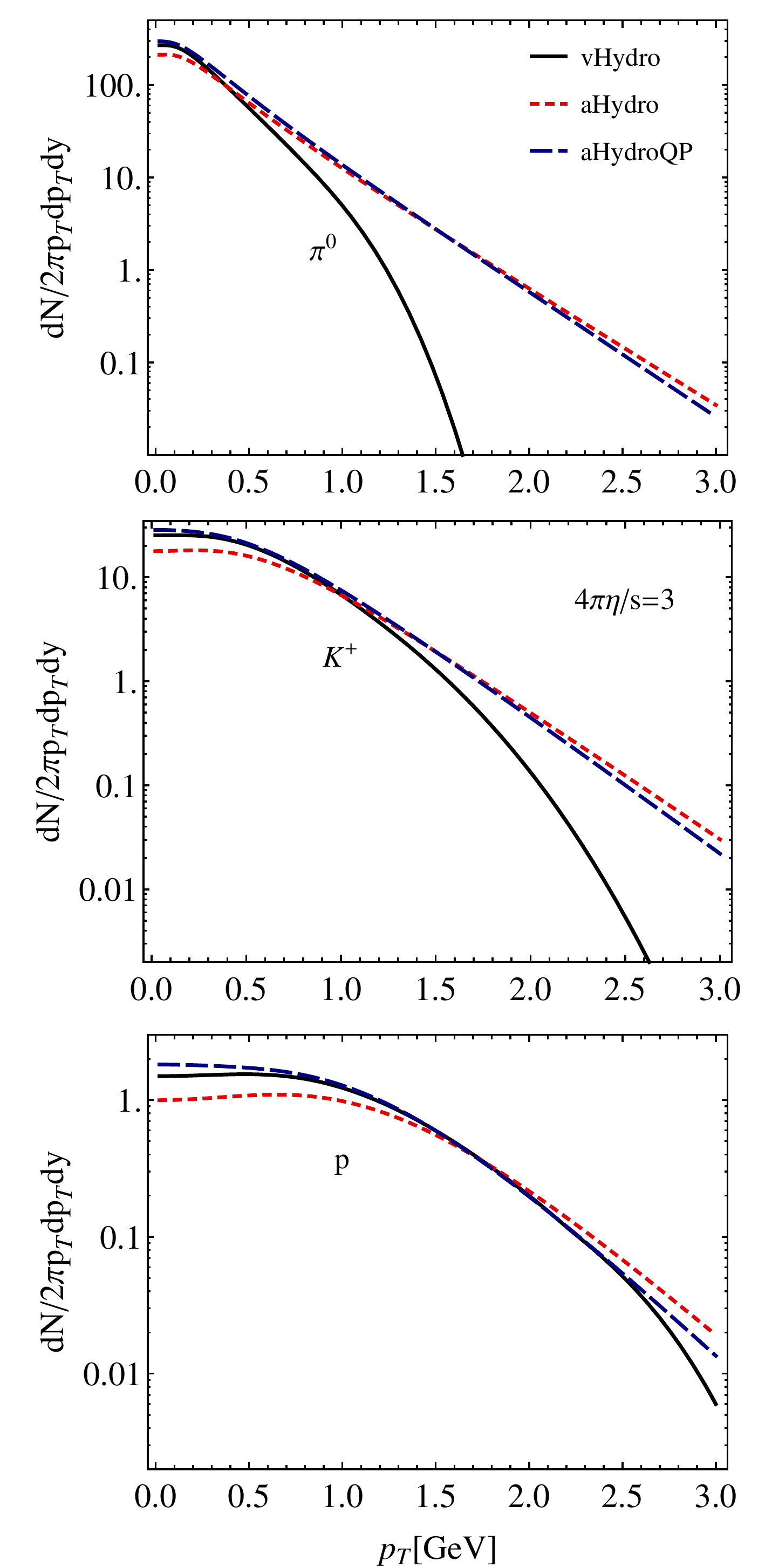}
\hspace{-10.7cm}
\includegraphics[width=13pc]{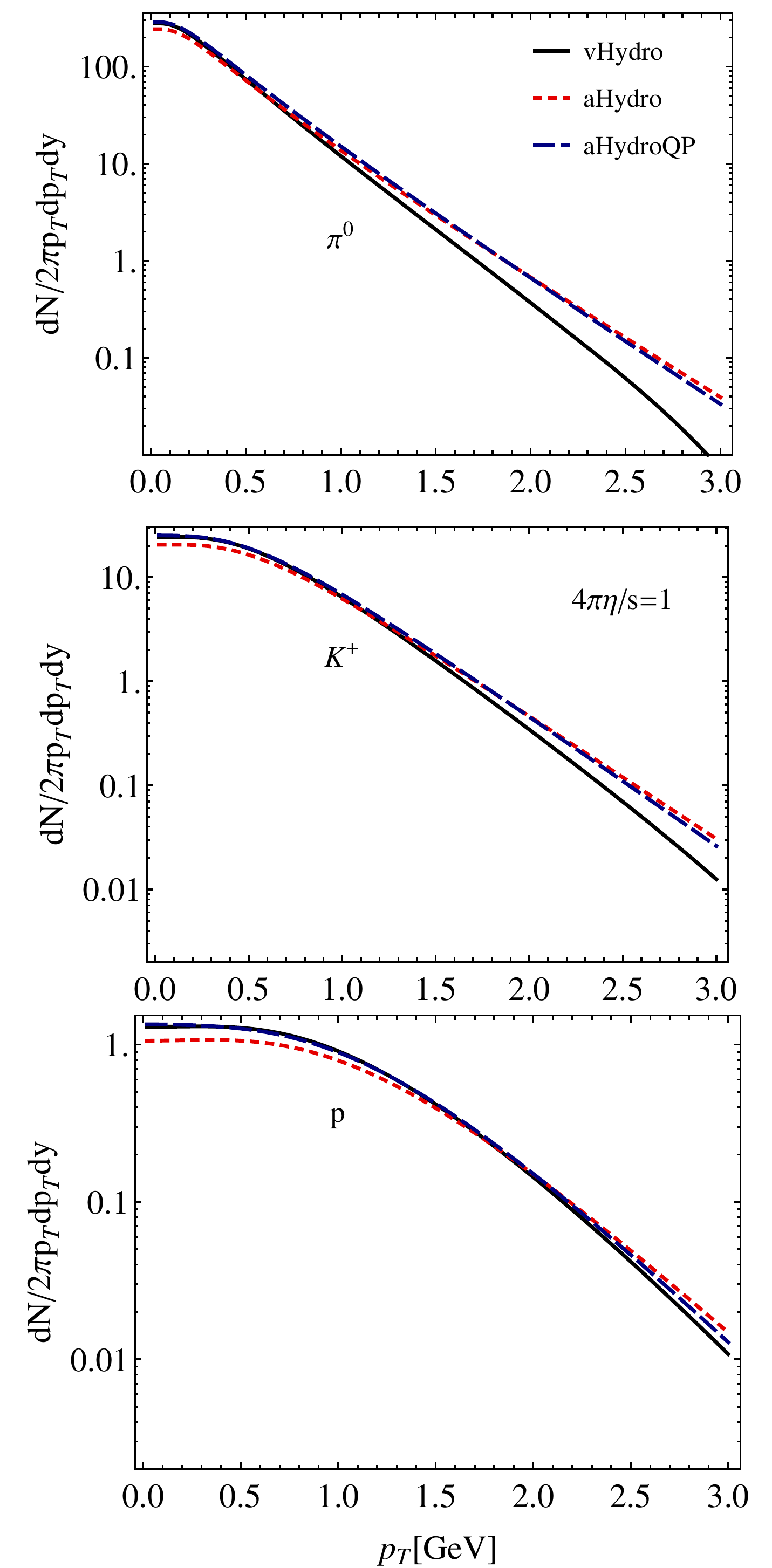}
\hspace{4.5cm}
\includegraphics[width=13pc]{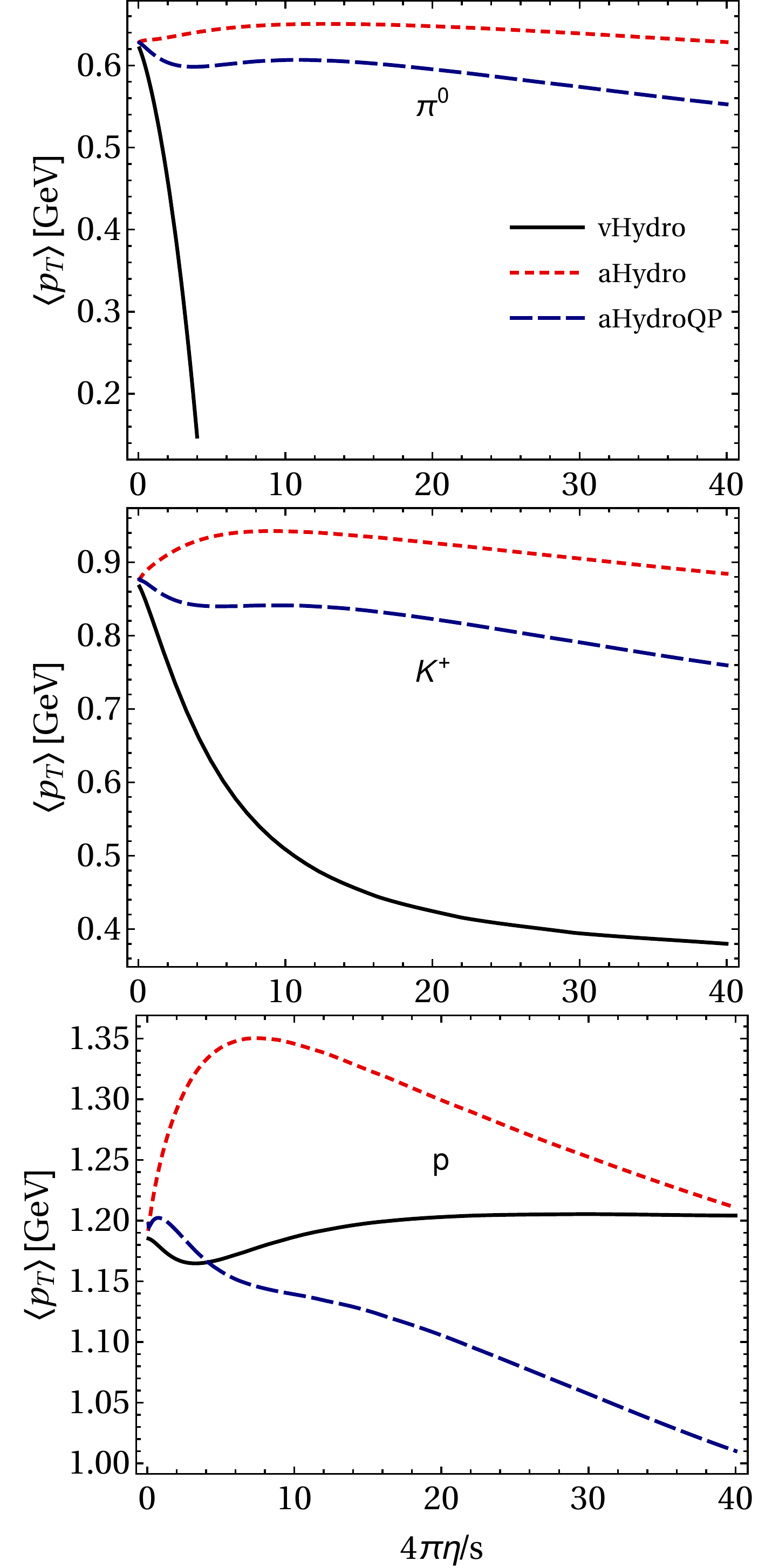}
\caption{The left column shows the spectra of neutral pions, kaons $(K^+)$, and
protons as a function of transverse momentum $ p_T$  with $ 4 \pi \eta / s =1$. The middle column is the same as the left column with $ 4 \pi \eta / s =3$. The right column shows a comparison of the neutral pion, kaon $(K^+)$, and
proton average transverse momentum as a function of $ 4 \pi \eta / s $. In all figures, the solid black line is the result from vHydro, the red dashed line is the result from aHydro, and the blue long-dashed line is the result from aHydroQP. Figures taken from Ref. \cite{Alqahtani:2016rth}.}
\label{fig:results}
\end{figure}

\section{Results}
We now turn to the numerical results. We present comparisons of the results obtained using anisotropic hydrodynamics (both aHydroQP and standard aHydro) and second-order viscous hydrodynamics. For all results presented here, we consider a central collision with an optical Glauber profile. We assume an  initial temperature of $T_0 = 600$ MeV at $\tau_0 = 0.25$ fm/c and take the freeze-out temperature to be $T_{\rm eff} = T_{\rm FO} = $ 150 MeV in all cases shown. In Fig.~\ref{fig:results} (left and middle columns), we present our results for the primordial pion, kaon, and proton spectra obtained for $4\pi\eta/s = 1$ and $3$, respectively. For $4\pi\eta/s=1$, both aHydro approaches are in good agreement over the entire $p_T$ range with the largest differences seen at low momentum.  The vHydro result, however, shows a significant downward curvature in the pion spectrum resulting in many fewer high-$p_T$ pions. For larger $\eta/s$ the vHydro primordial particle spectra become unphysical at lower momenta. For example, for $4\pi\eta/s=3$, the middle column, the differential pion spectrum goes negative at $p_T \sim 1.6$ GeV.

In Fig.~\ref{fig:results} (right column), we present the mean transverse momentum for pions, kaons, and protons as a function of $4\pi\eta/s $. We see that both aHydro approaches show a weak dependence of the pion and kaon $\langle p_T \rangle$ on the value of $\eta/s$. On the other hand, vHydro predicts a much more steep decrease in $\langle p_T \rangle$ for the pions and kaons. We  also see in vHydro that $\langle p_T \rangle$ for pions becomes negative for $4\pi\eta/s \sim 5$.  This rapid decrease can be understood from the negativity of the $ p_T$-differential pion spectra at high $p_T$ where the integrand is more sensitive to the high-$p_T$ part of the spectra.  Finally, we note that the proton spectra and hence $\langle p_T \rangle$ for protons is sensitive to the way in which the EoS is implemented when comparing the two aHydro approaches.

\section{Conclusions and outlook}
We compared three different viscous hydrodynamics approaches: aHydroQP, aHydro, and vHydro. We presented the predictions of these models for the primordial particle spectra and average transverse momentum. We found that they agree well for small shear viscosity to entropy density, $ \eta/s$ with the standard aHydro method showing suppressed production at low transverse-momentum compared to the other two methods. Finally, we showed that, when using standard viscous hydrodynamics, the bulk-viscous correction can drive the primordial particle spectra negative at large $p_T$, which is not seen with either aHydro method. 
 
Looking to the future, we will try to extend the aHydroQP method to 3+1d since this approach takes into account the non-conformality of the system from the beginning. However, this will be numerically intensive compared with the standard aHydro.  Until a 3+1d code is available, 1+1d results will be a reference point to show the differences between the two approaches used to impose a realistic EoS in aHydro.

%\section{}

\ack
We thank M. Nopoush for his collaboration. M. Strickland was supported by the U.S. Department of Energy under Award No.~\protect{DE-SC0013470}. M. Alqahtani was supported by a PhD fellowship from the University of Dammam.
 
%M. Alqahtani would like to thank the organizers of Hot Quarks 2016 conference for their help.and by Kent State University GSS travel award

\section*{References}

\bibliography{HQ16}

\end{document}